\documentclass{IEEEtran}
\usepackage{lettrine}
\usepackage{amsmath}
\usepackage{cases}
\usepackage{bm}
\usepackage{txfonts}
\usepackage{amssymb}
\usepackage{cite}
\usepackage{graphicx}
\usepackage{psfrag}
\usepackage{url}
\usepackage{stfloats}
\usepackage{algorithm}
\usepackage{algorithmic}        
\usepackage{multirow}
\usepackage{color}
\usepackage{romannum}
\usepackage{lipsum}
\usepackage{stfloats}
\usepackage[utf8]{inputenc}
\usepackage[skip=2pt,font=scriptsize]{caption}
\usepackage[skip=2pt,font=scriptsize]{subcaption}
\usepackage{tabularx}
\usepackage{booktabs}
\usepackage{eso-pic,xcolor}
\usepackage{overpic, scalerel}

\usepackage{dsfont}
\usepackage{minibox}
\DeclareSymbolFont{matha}{OML}{txmi}{m}{it}
\DeclareMathSymbol{\varv}{\mathord}{matha}{118}

\begin{document}
\pagenumbering{arabic}
\title{QoE-Aware and Secure UAV-Aided Rate-Splitting Multiple Access Based Communications}
\author{Abuzar B. M. Adam,~\IEEEmembership{Member,~IEEE}, Xiaoyu Wan, Mohammed Saleh Ali Muthanna,~\IEEEmembership{Member,~IEEE}
\thanks{A. B. M. Adam is with the Interdisciplinary Centre for Security, Reliability and Trust (SnT), University of Luxembourg, 1855, Luxembourg City,
Luxembourg.}
\thanks{X. Wan, is with School of Communications and Information Engineering, Chongqing University of Posts and Telecommunications, Chongqing,
P. R. China, 400065.}
\thanks{M. S. A. Muthanna is with Institute of Computer Technologies and Information Security, Southern Federal University, 347922 Taganrog, Russia.}
}

\maketitle
\begin{abstract}
In this work, we address the issue of quality of experience (QoE) in unmanned aerial vehicle (UAV) aided multiuser rate-splitting multiple access (RSMA) networks under secrecy constraints. The problem is formulated as maximization of sum mean opinion scores (MOSs) of the users. The problem is decomposed into two subproblems, beamforming and rate allocation and UAV trajectory subproblem. For, beamforming and rate allocation subproblem, we use the epigraph method, property of polynomials, and the norm-bounded error of channels, we linearize the objective function. Then, applying second-order conic (SOC) and first Taylor expansion, we convexify the remaining nonconvex constraints. For the highly nonconvex UAV trajectory, we unroll the constraints and we apply first Taylor expansion on the unrolled constraints. The simulation results demonstrate the efficiency of the proposed framework.
\end{abstract}
\begin{IEEEkeywords}
Unmanned aerial vehicle (UAV), Rate-splitting multiple access (RSMA), beamforming, rate allocation, trajectory optimization.
\end{IEEEkeywords}

\section{Introduction}

\IEEEPARstart{T}{he} demand for high data rates and reliable connectivity is drastically increasing with the explosive growth of wireless networks. In such densely deployed networks, it is difficult to deliver the services to every user. Therefore, unmanned aerial vehicle (UAV) is considered to provide uninterrupted service to inaccessible users by the terrestrial networks \cite{bb1,bb2,abm0}. 

The integration of UAVs equipped with multiple antennas has demonstrated significant potential in enhancing the energy efficiency of wireless communication networks \cite{abm3}. This improvement is largely attributed to the advanced spatial diversity and beamforming capabilities provided by multiple antenna systems, which enable more efficient use of the radio spectrum and reduced energy consumption.

Despite these advantages, the inherent broadcast nature of UAV communication channels poses substantial security challenges. These channels are particularly susceptible to various forms of malicious attacks, including eavesdropping, jamming, and spoofing, which can severely compromise the integrity and confidentiality of the transmitted data \cite{abm2}.

Consequently, the design and implementation of efficient and secure communication networks are of paramount importance for the advancement of next-generation networks. This entails the development of robust encryption techniques, secure communication protocols, and advanced intrusion detection systems to mitigate potential security threats while maintaining high levels of performance and energy efficiency \cite{abm4,abm5}. Addressing these challenges is critical to realizing the full potential of UAV-assisted communication networks in future wireless systems.

UAV can be effectively deployed using two main strategies to optimize their performance in wireless communication networks. The first strategy is static deployment, which employs three-dimensional (3D) placement techniques to determine the best fixed position for the UAV. This static positioning aims to maximize network coverage and minimize interference, thereby enhancing overall network efficiency \cite{abm6}.

The second strategy involves dynamic deployment, where the UAV's 3D trajectory is optimized in real-time. This dynamic approach allows the UAV to continuously adjust its flight path in response to changing environmental conditions and network requirements. By optimizing the 3D trajectory, UAVs can maintain optimal coverage, improve connectivity, and adapt to varying user demands and interference levels \cite{abm7}.

Both static 3D placement and dynamic 3D trajectory optimization are essential for the efficient deployment of UAVs in next-generation wireless networks. These strategies ensure that UAVs can deliver reliable and high-quality communication services, whether in fixed positions or while in motion.

Rate-splitting multiple access (RSMA) is considered as a powerful technology to improve the spectral efficiency \cite{bb3} and better suitable paradigm for multiple antenna communication networks \cite{bb5}. Hence, recently, UAV-aided RSMA has been deployed to achieve better network performance \cite{bb6,abm1}.

Putting into considerations the different needs of different users in the network, different quality-of-experience (QoE) metrics and algorithms are developed to measure users' satisfaction \cite{bb7,bb8}. Hence, several studies investigated QoE optimization in UAV-aided networks \cite{bb9,bb10}. However, in addition to QoE, security issues arise. In order to jointly address these issues in UAV-aided RSMA communications, UAV trajectory and RSMA parameters need to be optimized under the subjective needs of each user. \par
In the literature, the issue of joint QoE and security in UAV-aided RSMA is not well investigated. Motivated by the above in addition to the possibility of improving the user satisfaction using QoE-aware and secure measurement can improve the system performance without additional cost, our contributions can be summarized as
\begin{itemize}
\item We propose a UAV-aided multiuser RSMA where we aim at maximizing QoE of users under secrecy constraints. The optimization problem is formulated as sum of MOSs of the users. We aim at solving the problem by jointly optimizing the beamforming, rate allocation, and UAV trajectory.
\item To solve the problem, we decompose the problem into beamforming and rate allocation subproblem and UAV trajectory subproblem. For the beamforming and rate allocation subproblem, we linearize the objective function by using the epigraph method and property of polynomials in \cite{bb12}, and the definition of the norm-bounded error of the channels. To convexify the remaining nonconvex constraints, we apply second-order conic (SOC) and first Taylor expansion.
\item The UAV trajectory is nonconvex, therefore, as a first step, we unroll the constraints. The second step is applying first Taylor expansion on the unrolled constraints.
\item The simulation results show the efficiency of the proposed framework and demonstrate.
\end{itemize}

The rest of the paper is organized as follows. In Section II, we introduce the system model and the problem formulation. In Section III, we discuss the proposed iterative solution for the problem. Section IV includes the simulation results. Finally, Section V concludes the paper.

\section{System Model and Problem Formulation}
We consider a downlink RSMA network consists of $L$ legitimate users denoted as ${\cal L} = \left\{ {1,2,\cdots,L} \right\}$  and $K$ eavesdroppers denoted as  ${\cal K} = \left\{ {1,2,\cdots,K} \right\}$. The network is assisted with a single UAV $u$ equipped with $N$ antennas to deliver contents. The locations of the UAV and the users are denoted as $\left( {{x_u},{y_u},{z_u}} \right)$ and $\left( {{x_i},{y_i},0} \right),$, respectively. Where $i \in \{l,k\}$ with  $l \in {\cal L},k \in {\cal K}$. The UAV flies at constant height $z_u$ with constant speed $v_u$. The UAV trajectory is defined as ${\mathbf{q}}\left[ t \right] = \left( {{x_u}\left[ t \right],{y_u}\left[ t \right],{z_u}} \right)$  with  $t \in {\cal T} = \left\{ {1,2,\cdots, T}\right\}$, ${{\mathbf{q}}_0}$, and ${{\mathbf{q}}_F}$ represent the time slot, the initial point, and the final point in the trajectory, respectively. The channels are modeled as independent identically distributed Rayleigh fading channels, where ${{\mathbf{h}}_{u,i}} \in {\mathbb{C}^{N \times 1}}$  is the channel coefficients between $u$ and $i$ and given as
\begin{align}\label{eq0}
{\mathbf{h}}_{u,i}[t]&=\sqrt{\frac{{P{L^{LoS}}d_{u,i}^{-\alpha}{\Gamma _{u,i}}}}{{1+{\Gamma _{u,i}}}}}{\mathbf{h}}_{u,i}^{LoS}[t]\nonumber\\ &+\sqrt{\frac{{P{L^{NLoS}}d_{u,i}^{-\alpha}}}{{1+{\Gamma _{u,i}}}}}{\mathbf{h}}_{u,i}^{NLoS}[t],
\end{align}
where ${P{L^{LoS}}}$ and ${P{L^{NLoS}}}$ are the path loss factors of line-of-sight (LoS) and non-line-of-sight (NLoS), $d_{u,i}$ is the distance between the user $i$ and the UAV $u$. ${\Gamma _{u,i}}$ is the Rician factor, ${\mathbf{h}}_{u,i}^{LoS}$ and ${\mathbf{h}}_{u,i}^{NLoS}$ are the LoS and NLoS components of ${\mathbf{h}}_{u,i}$, where the entries of ${\mathbf{h}}_{u,i}^{NLoS}$ are modeled as circularly symmetric complex Gaussian distribution with zero mean and variance 1.
In downlink RSMA, common message  $\mathbb{E}\left\{ {{{\left\| {{\mathbf{s}_0}} \right\|}^2}} \right\} = 1$ and user's private message $\mathbb{E}\left\{{{{\left\| {{\mathbf{s}_i}} \right\|}^2}} \right\} = 1$ are transmitted from UAV $u$. The received signal at the user $i$ is given as
\begin{equation}\label{eq1}
y_i[t]={\mathbf{h}}_{u,i}^H\left[ t \right]\left({\sum\nolimits_{i\in\left\{ {l,k}\right\}} {{{\mathbf{w}}_i}\left[ t \right]{s_i}}  + {{\mathbf{w}}_0}\left[ t \right]{s_0}} \right) + {z_i},
\end{equation}
where ${{\mathbf{w}}_i} \in {\mathbb{C}^{N \times 1}}$ is the precoding vector and ${z_i} \sim {\cal C}{\cal N}\left( {0,\sigma _i^2} \right)$ is the additive white Gaussian noise. The rate for decoding the common signal is given as
\begin{equation}\label{eq2}
R_0[t]={\log _2}\left({1+\frac{{{{\left| { {{\mathbf{h}}_{u,l}^H\left[ t \right]}{{\mathbf{w}}_0}\left[ t \right]} \right|}^2}}}{{\sum\limits_{l = 1}^L {{{\left| { { {\mathbf{h}}_{u,l}^H\left[ t \right]} {{\mathbf{w}}_l}\left[ t \right]} \right|}^2}}  + \sigma _l^2}}} \right),
\end{equation}
The decoding of the common message should be done by all legitimate users. To guarantee that, the rate of common message is set as  $\mathop {\min }\limits_l {R_0}\left[ t \right]$. Therefore, the constraint $\sum\limits_{l = 1}^L {{a_l}}  \le \mathop {\min }\limits_l {R_0}\left[ t \right]$ is set to the user $l$ to decode the common message. The achievable rate of the user $l$ to decode the private signal is given as \begin{equation}\label{eq3}
{R_l}\left[ t \right] = {\log _2}\left({1+\frac{{{{\left| { {{\mathbf{h}}_{u,l}^H\left[ t \right]}{{\mathbf{w}}_l}\left[ t \right]} \right|}^2}}}{{\sum\limits_{j = 1, j\neq l}^L {{{\left| { { {\mathbf{h}}_{u,l}^H\left[ t \right]} {{\mathbf{w}}_j}\left[ t \right]} \right|}^2}}  + \sigma _l^2}}} \right),
\end{equation}
The achievable rate of the eavesdropper $k$ to decode the private signal of user $l$ is given as
\begin{align}\label{eq4}
& R_k[t] =  \log_2\Biggl( 1 + \nonumber\\ &  \frac{{{{\left| {{\mathbf{h}}_{u,k}^H\left[ t \right]{{\mathbf{w}}_l}\left[ t \right]} \right|}^2}}}{{\sum\limits_{j = 1,j \ne l}^L {{{\left| {{\mathbf{h}}_{u,k}^H\left[ t \right]{{\mathbf{w}}_j}\left[ t \right]} \right|}^2}}  + {{\left| {{\mathbf{h}}_{u,k}^H\left[ t \right]{{\mathbf{w}}_0}\left[ t \right]} \right|}^2} + \sigma _l^2}} \Biggr)
\end{align}
The secrecy rate at the user $l$ is expressed as follows
\begin{equation}\label{eq5}
R_l^{\sec }\left[ t \right] = {\left[ {R_l^{tot}\left[ t \right] - \mathop {\max }\limits_k R_k^{tot}\left[ t \right]} \right]^ + },
\end{equation}
where $R_l^{tot}\left[ t \right] = {a_l}\left[ t \right] + {R_l}\left[ t \right]$, and $R_k^{tot}\left[ t \right] = {a_l}\left[ t \right] + {R_k}\left[ t \right]$. We use mean opinion score (MOS) as a utility function which is defined as
${\rm{MO}}{{\rm{S}}_l}\left[ t \right] = {\lambda _1}\ln \left( {R_l^{\sec }\left[ t \right]} \right) + {\lambda _l}$ and ${\lambda _l} = {\lambda _2} + {\lambda _1}\ln \left( {{W_l} \cdot \frac{1}{\Omega}} \right) \label{eq6}$,
where ${\lambda _1}$ and  ${\lambda _2}$ are constants that are set to 1.12 and 4.675 respectively. $\Omega$ is the size of the data. The optimization problem can be defined in \eqref{eq8:main}. Constraint \eqref{eq8:b} is to denote the power budget. Constraint \eqref{eq8:d} is guarantee the decoding of the common message. Constraint \eqref{eq8:e} is the secrecy constraints and constraints \eqref{eq8:f}, \eqref{eq8:g}, and \eqref{eq8:h} are for UAV movement. Problem \eqref{eq8:main} is nonconvex with respect to ${\mathbf{w}},a,$ and ${\mathbf{Q}}$. In the following section, we discuss the proposed solution to problem \eqref{eq8:main}.
\begin{subequations}\label{eq8:main}
\begin{align}
& \mathop {\max }\limits_{{\mathbf{w}},a,{\mathbf{Q}}}
 \sum\limits_{l = 1}^L {{\rm{MO}}{{\rm{S}}_l}\left[ t \right]}  \tag{\ref{eq8:main}} \\
& \text{s.t.} \hspace{0.5cm} {{\mathbf{w}}_l} \ge 0,\forall l,\label{eq8:a} \\
&{\left\| {{{\mathbf{w}}_0}\left[ t \right]} \right\|^2} + \sum\limits_{l = 1}^L {{{\left\| {{{\mathbf{w}}_l}\left[ t \right]} \right\|}^2}}  \le {P_{\max }},\label{eq8:b}\\
&{a_l}\left[ t \right] \ge 0, \forall l,\label{eq8:c}\\
&\sum\limits_{l = 1}^L {{a_l}\left[ t \right]}  \le {R_0},\forall l,\label{eq8:d}\\
&R_l^{tot}\left[ t \right] - \mathop {\max }\limits_{{{\mathbf{e}}_{u,k}}} R_k^{tot}\left[ l \right] \ge \eta ,\forall k,\label{eq8:e}\\
&{\left\| {{\mathbf{q}}\left[ T \right] - {{\mathbf{q}}_F}} \right\|^2} \le {D^2},\label{eq8:f}\\
&{\left\| {{\mathbf{q}}\left[ {t + 1} \right] - {\mathbf{q}}\left[ t \right]} \right\|^2} \le {D^2},\label{eq8:g}\\
&{\mathbf{q}}\left[ 1 \right] - {{\mathbf{q}}_0},\label{eq8:h}
\end{align}
\end{subequations}
\section{Alternative Optimization Method}
To handle the problem in \eqref{eq8:main}, we divide it into beamforming and rate allocation subproblem and UAV trajectory subproblem and we jointly optimize the two subproblems.
\subsection{Beamforming and Rate Allocation }
Fixing the UAV trajectory, beamforming and rate allocation is given as
\begin{subequations}\label{eq9:main}
\begin{align}
& \mathop {\max }\limits_{{\mathbf{w}},a} \sum\limits_{l = 1}^L {{\rm{MO}}{{\rm{S}}_l}\left[ t \right]}  \tag{\ref{eq9:main}}; \hspace{0.5cm}\text{s.t.} \hspace{0.5cm} \eqref{eq8:a}-\eqref{eq8:e},\notag
\end{align}
\end{subequations}
Using the epigraph method in \cite{bb12}, problem  \eqref{eq9:main} can be reformulated as
\begin{subequations}\label{eq11:main}
\begin{align}
& \mathop {\max }\limits_{{\mathbf{w}},a} \sum\limits_{l = 1}^L {{\lambda _1}\ln \left( {R_l^{\sec }\left[ t \right]} \right)} \text{s.t.}\eqref{eq8:a}-\eqref{eq8:e} \tag{\ref{eq11:main}} \\
& \& R_l^{\sec }\left[ t \right]\leq \left[ {{a_l} + R_l^{tot}\left[ t \right]-\mathop{\max }\limits_{{{\mathbf{e}}_{u,k}}} \left( {{a_l}+R_k^{tot}[t]}\right)}\right]\nonumber
\end{align}
\end{subequations}
where ${{\mathbf{e}}_{u,k}}$ is error due to the channel uncertainties. Since, eavesdropper's channel has uncertainties; therefore, the numerator of $R_k\left[ t \right]$ can be rewritten as
\begin{align}
\label{eq12}
&{\left| {{\mathbf{h}}_{u,k}^H \left[ t \right]{{\mathbf{w}}_l}\left[ t \right]} \right|^2}  \nonumber\\
& ={\mathbf{w}}_l^H\left[ t \right]\left(\begin{array}{l}
{{{\mathbf{\tilde h}}}_{u,k}}\left[ t \right]{\mathbf{\tilde h}}_{u,k}^H\left[ t \right] + {{{\mathbf{\tilde h}}}_{u,k}}\left[ t \right]{\mathbf{e}}_{u,k}^H\left[ t \right]\nonumber\\
 + {{\mathbf{e}}_{u,k}}\left[ t \right]{\mathbf{\tilde h}}_{u,k}^H\left[ t \right] + {{\mathbf{e}}_{u,k}}\left[ t \right]{\mathbf{e}}_{u,k}^H\left[ t \right]
\end{array} \right){{\mathbf{w}}_l}\left[ t \right]\\
 &= {\mathbf{w}}_l^H\left[ t \right]\left( {{{{\mathbf{\tilde H}}}_{u,k}}\left[ t \right] + \Delta {{\mathbf{h}}_{u,k}}\left[ t \right]} \right){{\mathbf{w}}_l}\left[ t \right]
\end{align}
where ${{\mathbf{\tilde H}}_{u,k}}\left[ t \right] = {{\mathbf{\tilde h}}_{u,k}}\left[ t \right]{\mathbf{\tilde h}}_{u,k}^H\left[ t \right]$ and the error matrix $\Delta {{\mathbf{h}}_{u,k}}\left[ t \right]$ is norm-bounded as follows
\begin{align}
\label{eq13}
\left\| {\Delta {{\mathbf{h}}_{u,k}}\left[ t \right]} \right\| & \le \left\| {{{{\mathbf{\tilde h}}}_{u,k}}\left[ t \right]{\mathbf{e}}_{u,k}^H\left[ t \right]} \right\| + \left\| {{{\mathbf{e}}_{u,k}}\left[ t \right]{\mathbf{\tilde h}}_{u,k}^H\left[ t \right]} \right\|  \nonumber\\
 & + \left\| {{{\mathbf{e}}_{u,k}}\left[ t \right]{\mathbf{e}}_{u,k}^H\left[ t \right]} \right\|
\nonumber\\
 &\le \left\| {{{{\mathbf{\tilde h}}}_{u,k}}\left[ t \right]} \right\|\left\| {{\mathbf{e}}_{u,k}^H\left[ t \right]} \right\| + \left\| {{{\mathbf{e}}_{u,k}}\left[ t \right]} \right\|\left\| {{\mathbf{\tilde h}}_{u,k}^H\left[ t \right]} \right\| \nonumber\\
 & + \left\| {{{\mathbf{e}}_{u,k}}\left[ t \right]} \right\|\varepsilon  \le {\varepsilon ^2} + 2\varepsilon \left\| {{{{\mathbf{\tilde h}}}_{u,k}}\left[ t \right]} \right\| = \upsilon
\end{align}
The estimated CSI can be expressed as ${{\mathbf{H}}_{u,k}}\left[ t \right] = {{\mathbf{\tilde H}}_{u,k}}\left[ t \right] + \Delta {{\mathbf{h}}_{u,k}}\left[ t \right]$. ${R_l}\left[ t \right]$ and ${R_k}\left[ t \right]$ can be rewritten as
\begin{equation}\label{eq14}
{R_l}\left[ t \right] = {\log _2}\left( {1 + \frac{{{\rm Tr}\left( {{{\mathbf{H}}_{u,l}}{{\mathbf{W}}_l}} \right)}}{{\sum\limits_{j = 1,j \ne l}^L {{\rm Tr}\left( {{{\mathbf{H}}_{u,l}}{{\mathbf{W}}_j}} \right)}  + \sigma _l^2}}} \right)
\end{equation}
\begin{equation}\label{eq15}
{R_k}[t]=\log _2\left[{1 + \frac{{{\rm Tr}\left( {{{\mathbf{H}}_{u,k}}{{\mathbf{W}}_l}{\mathbf{H}}_{u,k}^H} \right)}}{{{\rm Tr}\left( {{{\mathbf{H}}_{u,k}}\left( {{{\mathbf{W}}_0} + \sum\limits_{j = 1,j \ne l}^L {{{\mathbf{W}}_j}} } \right)} \right) + \sigma _l^2}}}\right]
\end{equation}
where ${{\mathbf{W}}_l} = {{\mathbf{w}}_l}\left[ t \right]{\mathbf{w}}_l^H\left[ t \right],{{\mathbf{W}}_k} = {{\mathbf{w}}_k}\left[ t \right]{\mathbf{w}}_k^H\left[ t \right]$ and ${{\mathbf{H}}_{u,l}} = {{\mathbf{h}}_{u,l}}\left[ t \right]{\mathbf{h}}_{u,l}^H\left[ t \right]$. Using the definition of the norm-bounded error, we can define the following lower and upper bounds
\begin{equation}\label{eq16}
\mathop {\max }\limits_{{{\mathbf{e}}_{u,k}}} {\rm Tr}\left( {{{\mathbf{H}}_{u,k}}{{\mathbf{W}}_l}} \right) = {\rm Tr}\left( {\left( {{{{\mathbf{\tilde H}}}_{u,k}}\left[ t \right] + \upsilon {\mathbf{I}}} \right){{\mathbf{W}}_l}} \right),
\end{equation}
\begin{equation}\label{eq17}
\mathop {\min }\limits_{{{\mathbf{e}}_{u,k}}} {\rm Tr}\left( {{{\mathbf{H}}_{u,k}}{{\mathbf{W}}_l}} \right) = {\rm Tr}\left( {\left( {{{{\mathbf{\tilde H}}}_{u,k}}\left[ t \right] + \upsilon {\mathbf{I}}} \right){{\mathbf{W}}_l}} \right),
\end{equation}
Let  $\beta  = {2^\eta }$, constraints \eqref{eq8:e} can be rewritten as
\begin{equation}\label{eq18}
\Upsilon \Lambda  \ge \frac{\beta }{{\beta  - 1}}{\rm Tr}\left( {{{\mathbf{H}}_{u,l}}{{\mathbf{W}}_l}} \right){\rm Tr}\left( {{{\mathbf{H}}_{u,k}}{{\mathbf{W}}_l}} \right)
\end{equation}
where $\Lambda  = {\rm Tr}\left( {{{\mathbf{H}}_{u,k}}\left( {{{\mathbf{W}}_0} + \sum\limits_{j = 1,j \ne l}^L {{{\mathbf{W}}_j}} } \right)} \right) + \sigma _l^2 + \frac{\beta }{{\beta  - 1}}{\rm Tr}\left( {{{\mathbf{H}}_{u,k}}{{\mathbf{W}}_l}} \right)$ and $\Upsilon  = {\rm Tr}\left( {{{\mathbf{H}}_{u,l}}{{\mathbf{W}}_l}} \right) + \left( {1 - \beta } \right)\left( {\sum\limits_{j = 1,j \ne l}^L {{\rm Tr}\left( {{{\mathbf{H}}_{u,l}}{{\mathbf{W}}_j}} \right)}  + \sigma _l^2} \right)$,
Using second order cone (SOC), we can transform \eqref{eq18} in the following form
\begin{equation}\label{eq19}
{\left\| {\begin{array}{*{20}{c}}
{2\sqrt {\frac{\beta }{{\beta  - 1}}} {\rm Tr}\left( {{{\mathbf{H}}_k}{{\mathbf{W}}_l}} \right)}\\
{\Lambda  - \Upsilon }
\end{array}} \right\|_2} \le \Lambda  + \Upsilon,
\end{equation}
where  ${{\mathbf{H}}_k} = {{\mathbf{h}}_{u,l}}\left[ t \right]{\mathbf{\tilde h}}_{u,k}^H\left[ t \right]$. Substituting \eqref{eq16} and \eqref{eq17} into \eqref{eq19}, we get
\begin{equation}\label{eq20}
{\left\| {\begin{array}{*{20}{c}}
{2\sqrt {\frac{\beta }{{\beta  - 1}}} {\rm Tr}\left( {\left( {{{\mathbf{h}}_{u,l}}\left[ t \right]{\mathbf{\tilde h}}_{u,k}^H\left[ t \right] + \upsilon {\mathbf{I}}} \right){{\mathbf{W}}_l}} \right)}\\
{a - \Upsilon }
\end{array}} \right\|_2} \le b + \Upsilon
\end{equation}

The variables $a$ and $b$ are obtained by substituting the uncertainties in \eqref{eq16} and \eqref{eq17} into  $\Lambda$. The problem is then reformulated as
\begin{subequations}\label{eq21:main}
\begin{align}
& && \mathop {\max }\limits_{\scriptstyle{\rm{w}},a,{c_1},{c_2},\hfill\atop
\scriptstyle{c_3},{c_4},\pi \hfill}
\sum\limits_{l = 1}^L {{\lambda _1}\ln \left( {R_l^{\sec }\left[ t \right]} \right)}  &   & \tag{\ref{eq21:main}} \\
& &&\text{s.t.} {\rm Rank}\left( {{{\mathbf{W}}_l}} \right) = 1,{\rm Rank}\left( {{{\mathbf{W}}_0}} \right) = 1,\label{eq21:a}\\
&            &&{\rm Tr}\left( {{{\mathbf{W}}_0}} \right) + \sum\limits_{l = 1}^L {{\rm Tr}\left( {{{\mathbf{W}}_l}} \right)}  \le {P_{\max }},\label{eq21:b}\\
&            &&{a_l}\left[ t \right] \ge 0,\label{eq21:c}\\
&            &&{\rm Tr}\left( {{{\mathbf{H}}_{u,l}}{{\mathbf{W}}_0}} \right) - {2^{\left( {\sum\limits_{l = 1}^L {{a_l}}  - 1} \right)}}\nonumber\\ &            &&\left( {\sum\limits_{l = 1}^L {{\rm Tr}\left( {{{\mathbf{H}}_{u,l}}{{\mathbf{W}}_l}} \right)}  + \sigma _l^2} \right) \ge 0,\label{eq21:d}\\
&            &&{\rm Tr}\left( {{{\mathbf{H}}_{u,l}}{{\mathbf{W}}_l}} \right) + \sum\limits_{j = 1,j \ne l}^L {{\rm Tr}\left( {{{\mathbf{H}}_{u,l}}{{\mathbf{W}}_j}} \right)}  + \sigma _l^2 \ge {e^{{c_1}}},\label{eq21:e}\\
&            &&\sum\limits_{j = 1,j \ne l}^L {{\rm Tr}\left( {{{\mathbf{H}}_{u,l}}{{\mathbf{W}}_j}} \right)}  + \sigma _l^2 \le {e^{{c_2}}},\label{eq21:f}\\
&            &&
{\rm Tr}\left( {\left( {{{{\mathbf{\tilde H}}}_{u,k}}\left[ t \right] + \upsilon {\mathbf{I}}} \right){{\mathbf{W}}_l}} \right)
 + {\rm Tr}\Biggl( \left( {{{{\mathbf{\tilde H}}}_{u,k}}\left[ t \right] + \upsilon {\mathbf{I}}} \right) \nonumber\\ 
 &            &&\left( {{{\mathbf{W}}_0} + \sum\limits_{j = 1,j \ne l}^L {{{\mathbf{W}}_j}} } \right) \Biggl) + \sigma _l^2 \ge {e^{{c_3}}},\label{eq21:g}\\
&            &&{\rm Tr}\left( {\left( {{{{\mathbf{\tilde H}}}_{u,k}}\left[ t \right] + \upsilon {\mathbf{I}}} \right)\left( {{{\mathbf{W}}_0} + \sum\limits_{j = 1,j \ne l}^L {{{\mathbf{W}}_j}} } \right)} \right) + \nonumber\\ 
 &            &&\sigma _l^2 \le {e^{{c_4}}},\label{eq21:h}\\
&            &&{c_1} - {c_2} - {\pi _l} \ge \sum\limits_{l = 1}^L {{\lambda _1}\ln \left( {R_l^{\sec }\left[ t \right]} \right)} ,\label{eq21:i}\\
&            &&{c_3} - {c_4} \le {\pi _l},\label{eq21:j},
\end{align}
\end{subequations}
Constraints \eqref{eq21:a}, \eqref{eq21:f}, \eqref{eq21:i}, and \eqref{eq21:j} are nonconvex. We can circumvent the rank-one issue in \eqref{eq21:a} by approximating the beamforming vectors as follows
\begin{equation}\label{eq22}
\begin{array}{l}
{{\mathbf{W}}_l} = {\mathbf{w}}_l^{\left( \tau  \right)}{\mathbf{w}}_l^H + {{\mathbf{w}}_l}{\mathbf{w}}_l^{\left( \tau  \right),H}- {\mathbf{w}}_l^{\left( \tau  \right)}{\mathbf{w}}_l^{\left( \tau  \right),H},\\
{{\mathbf{W}}_0} = {\mathbf{w}}_0^{\left( \tau  \right)}{\mathbf{w}}_0^H + {{\mathbf{w}}_0}{\mathbf{w}}_0^{\left( \tau  \right),H} - {\mathbf{w}}_0^{\left( \tau  \right)}{\mathbf{w}}_0^{\left( \tau  \right),H},
\end{array}
\end{equation}
To address the issue in \eqref{eq21:f} and \eqref{eq21:i}; first, using auxiliary variable $\mathfrak{q}$, first Taylor expansion around  $\mathfrak{q}^{\left( \tau  \right)}$, and SOC, we get the following
\begin{equation}\label{eq23}
\begin{array}{l}
\frac{{{c_1} - {c_2} - {\pi _l} + 1}}{2} \ge {\left\| {{{\left[ {\frac{{{c_1} - {c_2} - {\pi _l} + 1}}{2},\mathfrak{q}} \right]}^T}} \right\|_2},\\
2{\mathfrak{q}^{\left( \tau  \right)}}\mathfrak{q} - {\left( {{\mathfrak{q}^{\left( \tau  \right)}}} \right)^2} \ge \sum\limits_{l = 1}^L {{\lambda _1}\ln \left( {R_l^{\sec }\left[ t \right]} \right)}
\end{array}
\end{equation}
Applying first Taylor expansion around $c_2^{\left( \tau  \right)}$ and $c_4^{\left( \tau  \right)}$ we transform \eqref{eq21:f} and \eqref{eq21:i} into the following
\begin{equation}\label{eq24}
\sum\limits_{j = 1,j \ne l}^L {{\rm Tr}\left( {{{\mathbf{H}}_{u,l}}{{\mathbf{W}}_j}} \right)}  + \sigma _l^2 \le {e^{c_2^{\left( \tau  \right)}}}\left( {{c_2} - c_2^{\left( \tau  \right)} + 1} \right),
\end{equation}
\begin{align}\label{eq25}
{\rm Tr}&\left( {\left( {{{{\mathbf{\tilde H}}}_{u,k}}\left[ t \right] + \upsilon {\mathbf{I}}} \right)\left( {{{\mathbf{W}}_0} + \sum\limits_{j = 1,j \ne l}^L {{{\mathbf{W}}_j}} } \right)} \right) + \nonumber\\ &\sigma _l^2 \le {e^{c_4^{\left( \tau  \right)}}}\left( {{c_4} - c_4^{\left( \tau  \right)} + 1} \right),
\end{align}

The final optimization problem is given as
\begin{subequations}\label{eq26:main}
\begin{align}
& \mathop {\max }\limits_{\scriptstyle{\rm{w}},a,{c_1},{c_2},\hfill\atop
\scriptstyle{c_3},{c_4},\pi ,\mathfrak{q}\hfill}
&& \sum\limits_{l = 1}^L {{\lambda _1}\ln \left( {R_l^{\sec }\left[ t \right]} \right)}  &   & \tag{\ref{eq26:main}} \\
& \text{s.t.}&& \eqref{eq21:a},\eqref{eq21:b}, \eqref{eq21:c}, \eqref{eq21:d}, \eqref{eq21:e}, \eqref{eq21:g}, \eqref{eq21:j},\notag\\
&            &&\eqref{eq20}, \eqref{eq23}, \eqref{eq24}, \eqref{eq25},\notag
\end{align}
\end{subequations}
The problem in \eqref{eq26:main} is convex and can be solved numerically using CVX.

\begin{algorithm}[H]
\caption {Proposed alternative optimization method (AOM) for solving \eqref{eq8:main}}
\begin{algorithmic}[1]
\renewcommand{\algorithmicrequire}{\textbf{Initialization:}}
\REQUIRE ${\mathbf{W}}^{\left(0\right)}$, ${\mathbf{Q}}^{\left(0\right)}$, ${\mathbf{a}}$, ${\mathbf{c}}^{\left(0\right)}= \{c_1^{\left(0\right)}, c_2^{\left(0\right)}, c_3^{\left(0\right)}, c_4^{\left(0\right)}, \mathfrak{q}^{\left(0\right)}\}$, $\Xi^{\left(0\right)}$, $\Omega$, ${{\rm MOS}_l^{\left( 0\right)}}$, $\epsilon$, $\tilde t=0$, and $\tau = 0$.
\WHILE{$\left| \sum\limits_{l=1}^{L}{{\rm MOS}_l^{\left(\tilde t\right)}} - \sum\limits_{l=1}^{L}{{\rm MOS}_l^{\left(\tilde t - 1\right)}}\right| > \epsilon $}
\STATE Solve problem \eqref{eq26:main} to obtain ${\mathbf{W}}^{\left(\tilde t\right)}$, ${\mathbf{a}}^{\left(\tilde t\right)}$, ${\mathbf{c}}^{\left(\tilde t\right)}$, $\Xi^{\left(\tilde t\right)}$
\STATE Solve \eqref{eq29:main} to obtain ${\mathbf{Q}}^{\left(\tilde t\right)}$
\STATE $t = t + 1$
\ENDWHILE
\end{algorithmic}
\end{algorithm}
\subsection{UAV Trajectory Optimization}
Given the rate allocation and the beamforming, the UAV trajectory can expressed as
\begin{subequations}\label{eq27:main}
\begin{align}
& \mathop {\max }\limits_{\mathbf{Q}}
&& \sum\limits_{l = 1}^L {{\lambda _1}\ln \left( {R_l^{\sec }\left[ t \right]} \right)}  &   & \tag{\ref{eq27:main}} \\
& \text{s.t.}&& \eqref{eq8:f},\eqref{eq8:g}, \eqref{eq8:h},\notag
\end{align}
\end{subequations}
We adopt the mechanism in \cite{bb13} to handle the nonconvex nature of the problem in \eqref{eq27:main}. The problem can be transformed in the following form
\begin{subequations}\label{eq28:main}
\begin{align}
& \mathop {\max }\limits_{{\rm{Q,}}\Xi }
&& \sum\limits_{l = 1}^L {{\lambda _1}\ln \left( {{\tilde R}_l^{\sec }\left[ t \right]} \right)}  &   & \tag{\ref{eq28:main}} \\
& \text{s.t.}&& \eqref{eq8:f},\eqref{eq8:g}, \eqref{eq8:h},\notag\\
&            &&\sqrt {{{\left( {{d_{u,l}}\left[ t \right]} \right)}^{ - \alpha }}}  \ge {\mathfrak{a}_l}\left[ t \right],\label{eq28:a}\\
&            &&\sqrt {{{\left( {{d_{u,k}}\left[ t \right]} \right)}^{ - \alpha }}}  \le {\mathfrak{b}_l}\left[ t \right],\label{eq28:b}\\
&            &&\sum\limits_{l = 1}^L {\tilde h_{u,l}^d\left[ t \right]{\mathbf{H}}_{u,l}^{\left( {\tilde t - 1} \right)}\tilde h_{u,l}^{d,T}\left[ t \right]{{\mathbf{W}}_l}}  \ge {\mathfrak{r}_{ll}}\left[ t \right],\label{eq28:c}\\
&            &&\sum\limits_{j = l + 1}^L {\tilde h_{u,l}^d\left[ t \right]{\mathbf{H}}_{u,l}^{\left( {\tilde t - 1} \right)}\tilde h_{u,l}^{d,T}\left[ t \right]{{\mathbf{W}}_j}}  + \sigma _l^2 \le {\mathfrak{r}_{lj}}\left[ t \right],\label{eq28:d}\\
&            &&\sum\limits_{k = 1}^K {\tilde h_{u,k}^d\left[ t \right]{\mathbf{H}}_{u,k}^{\left( {\tilde t - 1} \right)}\tilde h_{u,k}^{d,T}\left[ t \right]\left( {{{\mathbf{W}}_0} + \sum\limits_{l = 1}^L {{{\mathbf{W}}_l}} } \right)}  + \nonumber \\ 
&            &&\sigma _l^2 \le {\mathfrak{r}_{kk}}\left[ t \right],\label{eq28:e}\\
&            &&\tilde h_{u,k}^d\left[ t \right]{\mathbf{H}}_{u,k}^{\left( {\tilde t - 1} \right)}\tilde h_{u,k}^{d,T}\left[ t \right]\left( {{{\mathbf{W}}_0} + \sum\limits_{j = l + 1}^L {{{\mathbf{W}}_j}} } \right) + \nonumber \\ 
&            && \sigma _l^2 \le {\mathfrak{r}_{kj}}\left[ t \right],\label{eq28:f}\\
&            &&{\mathfrak{b}_l}\left[ t \right] \le \sqrt {{{\left( {{z_u}\left[ t \right]} \right)}^{ - \alpha }}} ,\label{eq28:g}
\end{align}
\end{subequations}

where ${{\tilde R}_l^{\sec }\left[ t \right]}$ is given on top of the next page.

\begin{figure*}
\begin{equation*}
\begin{aligned}
&\tilde R_l^{\sec}\left[t\right] = {\log _2}\left( {\sigma _l^2 + \sum\limits_{l = 1}^L {h_{u,l}^d\left[ t \right]{\mathbf{H}}_{u,l}^{\left( {\tilde t - 1} \right)}h_{u,l}^{d,T}\left[ t \right]{{\mathbf{W}}_l}} } \right) - {\log _2}\left( {\sum\limits_{j = l + 1}^L {h_{u,l}^d\left[ t \right]{\mathbf{H}}_{u,l}^{\left( {\tilde t - 1} \right)}h_{u,l}^{d,T}\left[ t \right]{{\mathbf{W}}_j}}  + \sigma _l^2} \right)\\
&- {\log _2}\left( {\sum\limits_{k = 1}^K {h_{u,k}^d\left[ t \right]{\mathbf{H}}_{u,k}^{\left( {\tilde t - 1} \right)}h_{u,k}^{d,T}\left[ t \right]\left( {{{\mathbf{W}}_0} + \sum\limits_{l = 1}^L {{{\mathbf{W}}_l}} } \right)}  + \sigma _l^2} \right) + {\log _2}\left( {h_{u,k}^d\left[ t \right]{\mathbf{H}}_{u,k}^{\left( {\tilde t - 1} \right)}h_{u,k}^{d,T}\left[ t \right]\left( {{{\mathbf{W}}_0} + \sum\limits_{j = l + 1}^L {{{\mathbf{W}}_j}} } \right) + \sigma _l^2} \right)\\
 &= {\log _2}\left( {1 + {\mathfrak{r}_{ll}}\left[ t \right]} \right) - {\log _2}\left( {1 + {\mathfrak{r}_{lj}}\left[ t \right]} \right)- {\log _2}\left( {1 + {\mathfrak{r}_{kk}}\left[ t \right]} \right) + {\log _2}\left( {1 + {\mathfrak{r}_{kj}}\left[ t \right]} \right),
\end{aligned}
\end{equation*}
\end{figure*}
$h_{u,l}^d\left[ t \right] = \left[ {\sqrt {{{\left( {{d_{u,l}}\left[ t \right]} \right)}^{ - \alpha }}} } \right],h_{u,k}^d\left[ t \right] = \left[ {\sqrt {{{\left( {{d_{u,k}}\left[ t \right]} \right)}^{ - \alpha }}} } \right]$,\\
$\tilde h_{u,l}^d\left[ t \right] = \left[ {{\mathfrak{a}_l}\left[ t \right]} \right],\tilde h_{u,k}^d\left[ t \right] = \left[ {{\mathfrak{b}_l}\left[ t \right]} \right]$,
$\left( {\tilde t - 1} \right)$ is the last iteration in the solution.  ${\mathbf{H}}_{u,l}^{\left( {\tilde t - 1} \right)} = {\mathbf{\tilde H}}_{u,k}^{\left( {\tilde t - 1} \right)}\left[ t \right] + \upsilon {\mathbf{I}}$. $\Xi  = \left\{ {\left\{ {{\mathfrak{a}_l}\left[ t \right]} \right\}_{l = 1}^L,\left\{ {{\mathfrak{b}_l}\left[ t \right]} \right\}_{l = 1}^L,\left\{ {{\mathfrak{r}_{lj}}\left[ t \right]} \right\}_{l = 1}^L,\left\{ {{\mathfrak{r}_{kk}}\left[ t \right]} \right\}_{k = 1}^K} \right\}$ is a set of slack variables. We unroll constraints \eqref{eq28:a} and \eqref{eq28:b} and applying first Taylor expansion on the nonconvex parts in the unrolled constraints as well as \eqref{eq28:c}, we can approximate problem \eqref{eq28:main} into
\begin{subequations}\label{eq29:main}
\begin{align}
& &&\mathop {\max }\limits_{{\mathbf{Q,}}\Xi }
\sum\limits_{l = 1}^L {{\lambda _1}\ln \left( \begin{array}{l}
{\log _2}\left( {1 + {\mathfrak{r}_{ll}}\left[ t \right]} \right) - \frac{{{\mathfrak{r}_{lj}}\left[ t \right]}}{{\ln 2\left( {1 + \mathfrak{r}_{lj}^{\left( \tau  \right)}\left[ t \right]} \right)}}\\
 - \frac{{{\mathfrak{r}_{kk}}\left[ t \right]}}{{\ln 2\left( {1 + \mathfrak{r}_{kk}^{\left( \tau  \right)}\left[ t \right]} \right)}} + {\log _2}\left( {1 + {\mathfrak{r}_{kj}}\left[ t \right]} \right)
\end{array} \right)}  &   & \tag{\ref{eq29:main}} \\
& &&\text{s.t.} \hspace{0.5cm} \eqref{eq8:f},\eqref{eq8:g}, \eqref{eq8:h}, \eqref{eq28:d}, \eqref{eq28:e}, \eqref{eq28:f}, \eqref{eq28:g},\notag\\
&  &&
x_u^2\left[ t \right] + x_l^2 + y_u^2\left[ t \right] + y_l^2 - 2{x_u}\left[ t \right]{x_l} - 2{y_u}\left[ t \right]{y_l} + z_u^2\left[ t \right]
\nonumber\\&  &&- \left( {1 + \frac{4}{\alpha }} \right){\left( {\mathfrak{a}_l^{\left( \tau  \right)}\left[ t \right]} \right)^{ - {\textstyle{4 \over \alpha }}}} + \frac{4}{\alpha }{\left( {\mathfrak{a}_l^{\left( \tau  \right)}\left[ t \right]} \right)^{ - {\textstyle{4 \over \alpha }} - 1}}\nonumber\\&  &&{\left( {{\mathfrak{a}_l}\left[ t \right]} \right)^{ - {\textstyle{4 \over \alpha }}}} \le 0,\label{eq29:a}\\
&  &&
{\left({{\mathfrak{b}_l}[t]}\right)^{-{\textstyle{4\over\alpha}}}}+\hat x_u^2[t]+y_u^{2,(\tau  )}[ t]-2x_u^{(\tau)}[t]{x_u}[t]-2y_u^{(\tau)}[t]{y_u}[t]\nonumber\\&  &&
-x_k^2-y_k^2-2{x_u}[t]{x_k}-2{y_u}[t]{y_k}-z_u^2[t]\le0,\label{eq29:b}\\
&  &&
{\mathfrak{r}_{ll}}\left[ t \right] + \sum\limits_{l = 1}^L {\tilde h_{u,l}^{d,\left( \tau  \right)}\left[ t \right]{\mathbf{H}}_{u,l}^{\left( {\tilde t - 1} \right)}\tilde h_{u,l}^{d,T,\left( \tau  \right)}\left[ t \right]\frac{{{{\mathbf{W}}_l}}}{{\sigma _l^2}}} 
\nonumber\\&  &&
 - 2{\mathop{\rm Re}\nolimits} \left\{ {\sum\limits_{l = 1}^L {\tilde h_{u,l}^{d,\left( \tau  \right)}\left[ t \right]{\mathbf{H}}_{u,l}^{\left( {\tilde t - 1} \right)}\tilde h_{u,l}^{d,T}\left[ t \right]\frac{{{{\mathbf{W}}_l}}}{{\sigma _l^2}}} } \right\} \le 0,\label{eq29:c}
\end{align}
\end{subequations}
Problem \eqref{eq29:main} is convex and can be efficiently solved using CVX. Algorithm 1 illustrates the steps of solving problem \eqref{eq8:main}.

\section{Simulation Results}

In this section, we present the simulation results. The default simulation parameters are shown in TABLE \ref{table} unless stated otherwise.

\begin{table}[!ht]
\caption{Simulation parameters}
\label{table}
\centering
{
\begin{tabular}{l l}
  \hline
  \textbf{Parameter} & \textbf{Value} \\
  \hline
  Users distribution scheme & Randomly uniform distribution \\
  Shadowing & Log-normal, standard deviation 8 dB \\
  Fading & Rayleigh fading with variance 1 \\
  Subcarrier frequency & 250 kHz\\
  Noise power spectral density & -174 dBm/Hz \\
  UAV antennas spacing & $\frac{\lambda}{2}$\\
  $v_u$ & 20 m/s\\
  $K$ & 2\\
  $\epsilon$ & 0.001\\
  $PL^{LoS}, PL^{NLoS}$ & -2.5 dB, -3.5 dB\\
  $\alpha$ & 3.5\\
  $\lambda$ & 0.15m\\
  $\Gamma_{u,i}$ & 2\\
  \hline
\end{tabular}
}
\end{table}

\begin{figure}[!ht]
\centerline{\includegraphics[width=2.5 in]{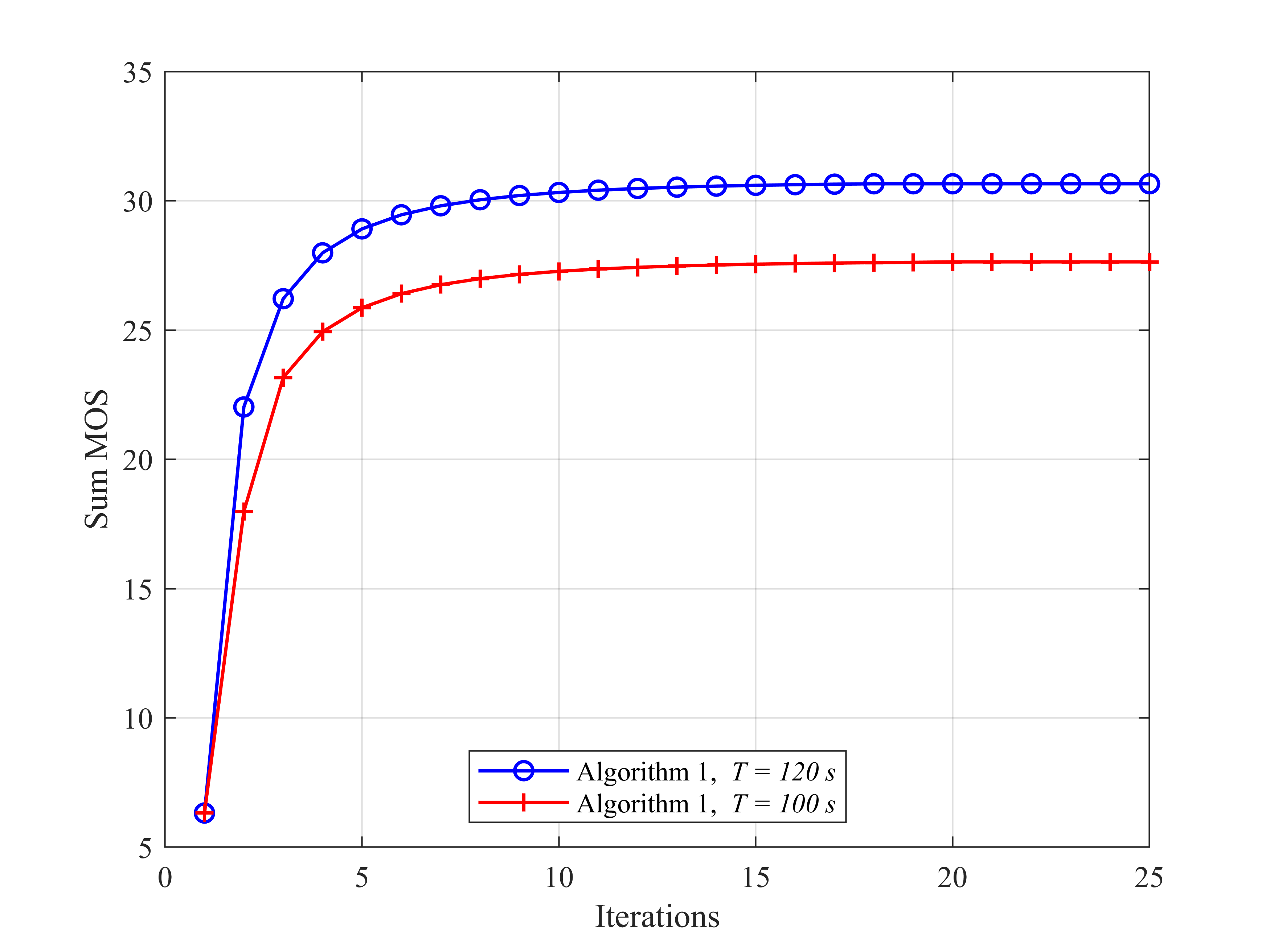}}
\caption{Convergence of Algorithm 1.}
\label{cv}
\end{figure}

From Fig. \ref{cv}, we can observe that Algorithm 1 exhibits a good convergence behavior in terms of sum MOS. Algorithm 1 converges in 20 iterations for both $T = 120$ s and $T = 100$ s. To plot Fig. \ref{cv}, the number of users is set to 4 and the other parameters are kept at their default values.

\begin{figure}[!ht]
\centering
\includegraphics[width=2.5in]{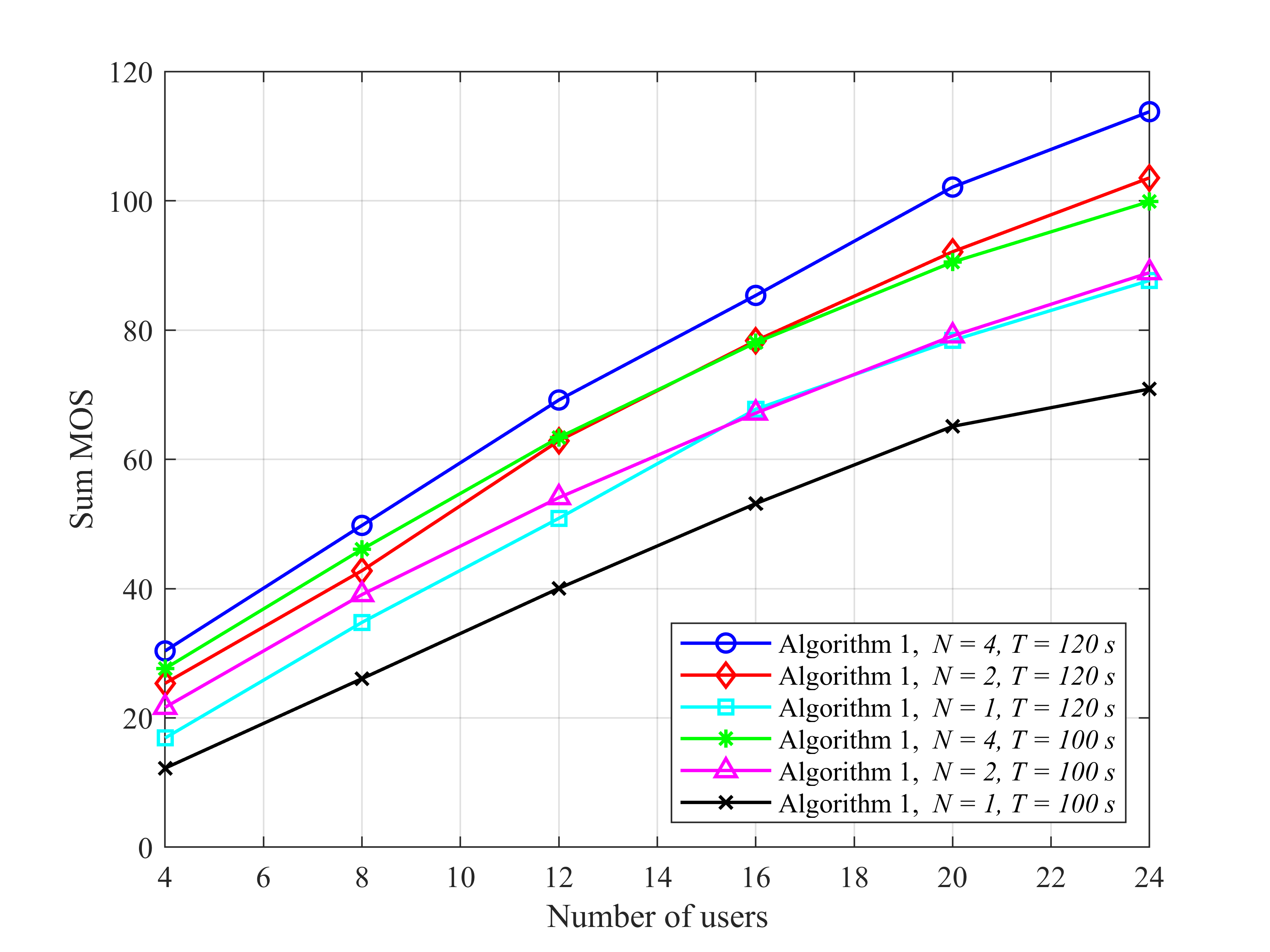}
\caption{Sum MOS for different number of user and $T$.}
\label{ue}
\end{figure}

Fig. \ref{ue} depicts the performance in terms of sum MOS for different number of users. Different values of $T$ are considered as well as number of antennas $N$. The longer the flying period $T$, the better the system performance. Also, it can be observed that when increasing $N$, the performance drastically improved.

\begin{figure}[!ht]
\centering
\includegraphics[width=2.5in]{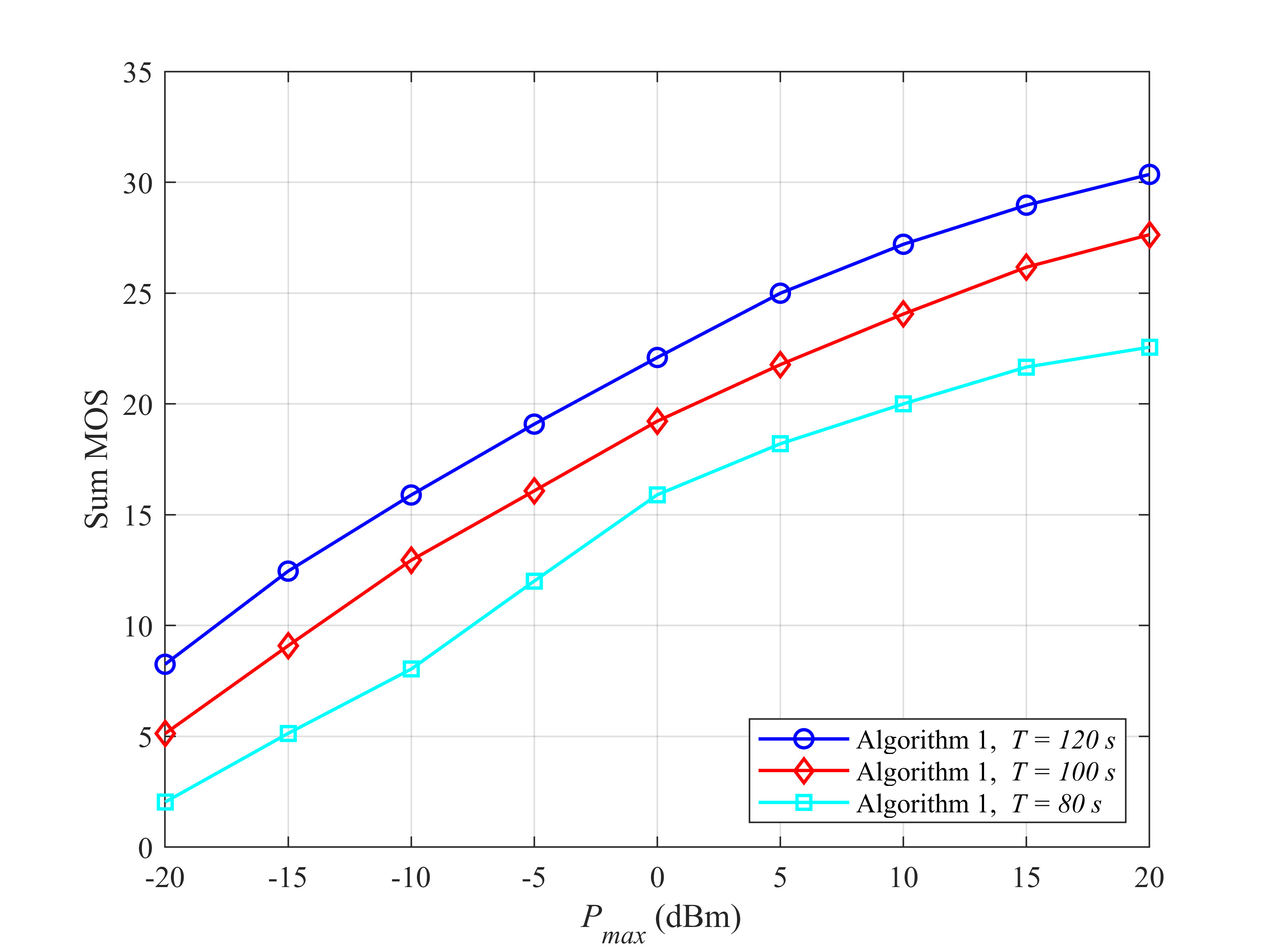}
\caption{Sum MOS for different values of $P_{max}$ and $T$.}
\label{bd}
\end{figure}

In Fig. \ref{bd}, we investigate the performance of the proposed framework for different values of maximum transmit power $P_{max}$. We fix the number of users to 4. From Fig. \ref{bd}, it is obvious that the increase in $P_{max}$ leads to a better performance in terms of sum MOS. This a trivial because from the definition of MOS in \eqref{eq6}, higher $P_{max}$ leads to higher data rate and by default leads to better MOS. Combined with the results in Fig. \ref{ue}, the flying period has a great impact on the performance.

\begin{figure}[!ht]
\centering
\includegraphics[width=2.5in]{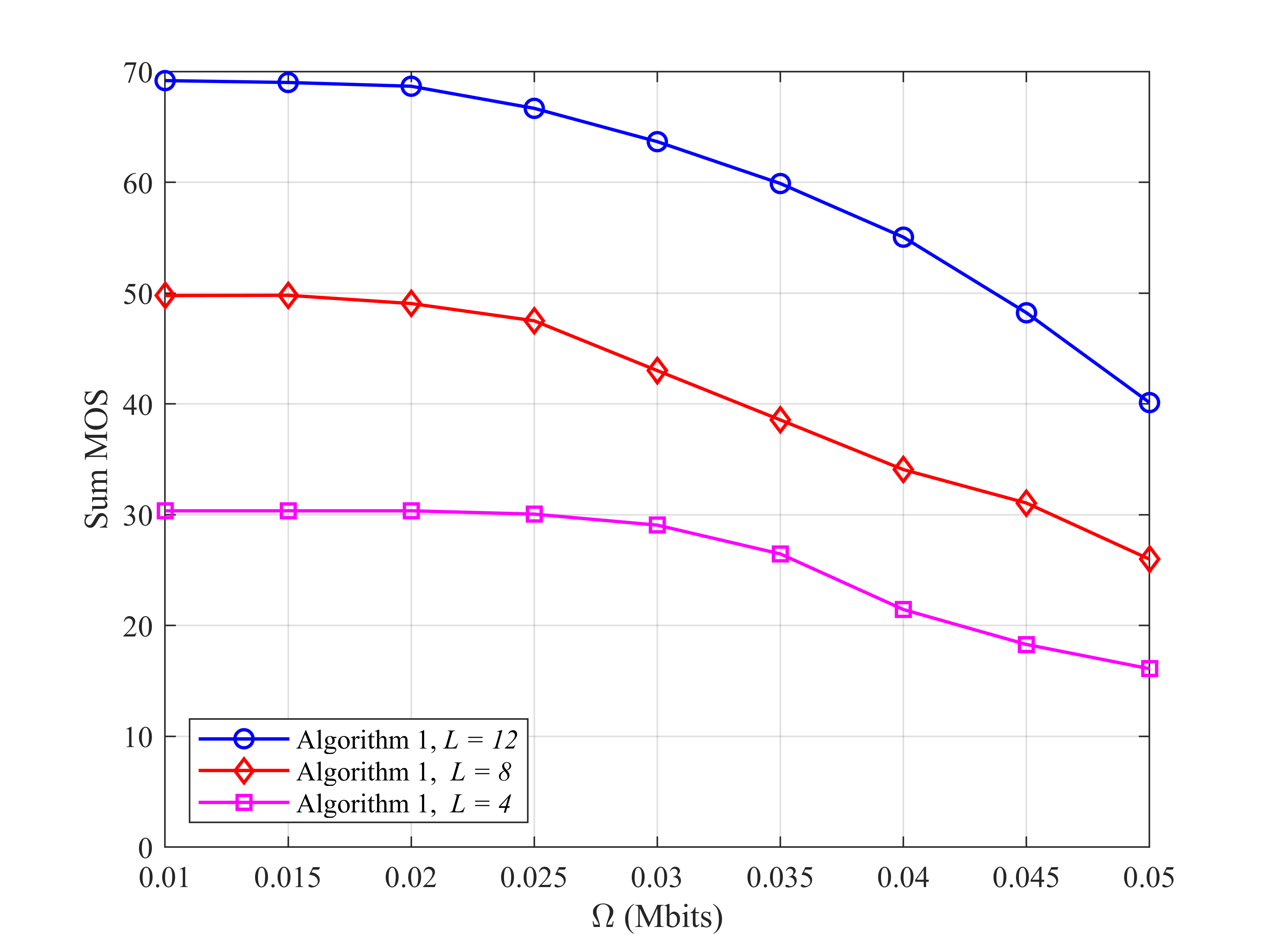}
\caption{Sum MOS for different values of $\Omega$ and $L$.}
\label{sz}
\end{figure}

Fig. \ref{sz} shows the change in sum MOS with different data sizes $\Omega$. It can be seen that with the increase in $\Omega$, the sum MOS decreases. This decreases proportionally with the number of users. For instance, when $L = 12$ sum MOS is drastically decreasing with the increase of $\Omega$. However, when $L = 4$,  the proposed framework is robust for $\Omega < 0.03$ Mbits, then it is gradually decreasing for higher values of $\Omega$.
\section{Conclusion}
In this work, we studied QoE maximization in in UAV-aided multiuser RSMA networks under the secrecy constraints. we formulated the problem as MOSs maximization then, we decoupled the problem into beamforming and rate allocation subproblem and UAV trajectory subproblem. For the first subproblem, we used the epigraph, property of polynomials, and the definition of the norm-bounded error of the channels to linearize the objective function. Then, SOC and first Taylor expansion were applied to convexify the nonconvex constraints. For the UAV trajectory, the constraints were unrolled and first Taylor expansion was applied to nonconvex unrolled constraints. The simulation results demonstrated the efficiency of the proposed framework.

\vfill
\end{document}